\documentclass[preprint,aps,draft]{revtex4}
\usepackage[dvips]{graphicx}
\usepackage[usenames]{color}
\usepackage{xspace,colortbl}
\begin{document}

\title{Contour-Variable Model of Constitutive Equations for Polymer Melts}
\author{Shuanhu Qi}
\author{Shuang Yang}
\author{Dadong Yan\footnote[1]{Corresponding author.
Tel./Fax: +86-10-82617358; {\it E-mail:} yandd@iccas.ac.cn.}}
\affiliation
{Beijing National Laboratory for Molecular Sciences (BNLMS), \\
Institute of Chemistry, Chinese Academy of Sciences, Beijing
100190, China}

\date{\today}

\baselineskip=24pt

\begin{abstract}
\bigskip
{Based on a modified expression of the rate of the convective
constraint release, we present a new contour-variable model of
constitutive equations in which the non-uniform segmental stretch
and the non-Gaussian chain statistical treatment of the single
chain are considered to describe the polymer chain dynamics and
the rheological behavior of an entangled system composed of linear
polymer chains. The constitutive equations are solved numerically
in the cases of steady shear and transient start-up of steady
shear. The results indicate that the orientation and stretch, as
well as the tube survival probability, have strong dependence on
the chain contour variable, especially in the high-shear-rate
region. However, the inclusion of the non-uniform features in the
constitutive models has little modification comparing with the
uniform models in determining the rheological properties both
qualitatively and quantitatively.}

\vspace{10mm} {\it Keywords:} Constitutive equation, CCR, Contour
variable, Polymer dynamics

\end{abstract}
\maketitle

\newpage

\section{INTRODUCTION}
Constitutive equations in polymer melts or concentrated solutions
are mathematical relationships between the stresses and the
external flow conditions. They represent the inherent properties
of the polymer system, and should be derived from the knowledge of
polymer chain structures, configurations, interactions and polymer
dynamics. In practice, a constitutive model is needed to idealize
these microscopic characteristics of polymer chains, due to the
complex interactions of this many-chain system (Larson, 1988). As
polymer melts own some characteristics of universality, that is to
say, some physical properties of polymer melts are independent on
the microscopic chemical structures, we can coarse grain a real
polymer chain as a smooth thread connected by segments (Watanabe,
1999). This coarse-grained chain, just like the real chain, has an
enormous number of configurations and can be usefully described
statistically as Gaussian chain or non-Gaussian chain. In fact,
the dynamics in our study is the segmental dynamics, which is
based on a spatial scale of a chain segment and those chains are
only distinguished by their lengths and abilities of extension.
The interactions between chains are intricate. However, the
attractive interactions between segments are tended to screen out
by the excluded volume effect in polymer melts or concentrated
solutions (de Gennes, 1979), leaving the topological constraints
to play a dominant role in such a system at equilibrium. A basic
understanding of chain dynamics is the key-point to develop
constitutive equations for melts or concentrated solutions of
flexible polymers. The proposition of the idea of `reptative'
motion by de Gennes (1971) is a milestone in the development of
polymer dynamics of the concentrated system. In fact, in a flow
the long linear polymer chains become oriented and stretched,
which generates internal stress as the response to external
disturbance. However, the orientation and stretch of chains will
relax in the course of time due to the motion of non-crosslinked
polymer chains. The major mechanisms of relaxation concerned are:
reptation (Doi and Edwards, 1978a, b, c, 1979), convection of
segments along primitive chains (Doi, 1980; Marrucci, 1986),
fluctuations of the contour length (Doi and Edwards, 1986; Pearson
and Helfand, 1984; Mead et al., 1998), reptative constraint
release (Marrucci, 1985; des Cloizeaux, 1988, 1990; Leygue et al.,
2005, 2006a, b), and convective constraint release (CCR)
(Marrucci, 1996; Ianniruberto and Marrucci, 1996; Milner et al.,
2001; Graham et al., 2003). The first three kinds of relaxation
are caused by the motion of the primitive chains, which can happen
only at the ends; while the other two are induced by the motion of
surrounding chains, which can occur not only at their ends but
also in some other place along the primitive chains.

The optical-stress law guarantees a linear relationship between
the stress and the orientational anisotropy of segments for a
highly crosslinked rubber system (Treloar, 1975). The subsequent
studies also show its validity for non-crosslinked polymer system
(Janeschitz-Kriegl, 1975). The segments in the crosslinked rubber
refer to chain portions between two permanently tethered crosslink
points. The orientations of these segments are uniform. The
segments in the non-crosslinked polymer system refer to
entanglement segments. The orientations of these segments are
non-uniform, as the segments can move freely. Under fast flow the
polymer chains not only become oriented but also significantly
stretched. The contribution from segmental stretch to stress is
proportional to the square of stretch ratio for Gaussian chains
(Doi and Edwards, 1986), while this relationship is much more
complex for non-Gaussian chains. In the original Doi and Edwards
(DE) model, due to the faster relaxation of segmental stretch, the
chain is assumed to remain its equilibrium length all the time.
The stress contribution comes only from segmental orientation. DE
theory predicts a maximum in shear stress $\sigma_{xy}$  with
respect to the shear rate, and followed by a sharp decrease
asymptotically as $\dot{ \gamma}^{-0.5}$, which has not been
observed by experiments. The neglect of segmental stretch results
in a monotonic increase in the first normal stress difference in
the start-up of a simple shear flow, however, an overshoot appears
in experiment. In order to improve this deficiency of the original
DE model, several `chain stretch' models are proposed. In the
uniform stretch model, it is assumed a uniform segmental stretch
along the chain, such as Larson's `partially convected strand'
model (1984) and also the one proposed by Pearson et al., (1989).
The non-uniform segmental stretch model is proposed by Marrucci
and Grizzuti (1988). The segmental stretch models are called as
DEMG model (Mead and Leal, 1995a, b). The extension of chain
segments creates some segment (or saying pseudo \emph{defect} (de
Gennes, 1971)) sources along the chain, which is equivalent to
reduce the diffusion coefficient. The inclusion of chain stretch
also changes the feature in the start-up of shearing. A serious
deficiency in all these models is that in the terminal region
excessive shear thinning is inevitable due to the lacking of an
extra relaxation mechanism to increase the rate of creating new
segments, which is equal to the rate of annihilating old segments,
or equivalently saying to avoid high orientation of the chain to
the direction of flow. The Mead-Larson-Doi (MLD) model
incorporating CCR to the DEMG model qualitatively improves the
description of the phenomenon of excessive shear thinning and
other rheological properties in steady state and transient
shearing flow (Mead et al., 1998). Recently Pattamaprom and Larson
(2001) proposed a toy MLD model by making an extension of the MLD
model in a simple way and adding reptative constraint release to
MLD equations. In all these models, the rate of CCR is evaluated
at the tube end, which is a uniform rate of CCR.

CCR represents the ability of the flow to convect or release
entanglements, and is determined by segmental orientation and
stretch. As the segmental orientation and stretch in the
non-crosslinked polymer melts are non-uniform, the CCR is
non-uniform. The rate of CCR is a function of both the contour
variable and time. We assume that when the entanglement is
convected away from the primitive chain segment to some other
place, the previously formed environment has changed and this
entanglement then becomes released. Based on this idea, we develop
a new contour-variable model including non-uniform segmental
stretch, non-uniform rate of CCR and non-Gaussian chain to
describe the rheological behavior of linear chain entangled system
in simple shearing and transient flows. However, we dropped the
reptative constraint release term in the present work as we focus
on the high-shear-rate region, where the segmental orientation and
stretch are obvious. In the low-shear-rate region, the reptative
constraint release occurs on a time scale of the reptation time of
the whole chain and one event only relaxes a small part of the
chain (Graham et al., 2003); while in the high-shear-rate region,
the CCR dominates. Thus in both of these two cases it is valid to
drop the reptative constraint release term. While in some middle
region, some quantitative deviations occur if we drop the
reptative constraint release term. We also neglected the
contribution from the fluctuations of the contour variable, as the
problem is considered in the mean-field level.

This paper is organized as follows. In section \ref{sec. 2} we
develop a contour-variable dependent rate of CCR to describe the
flow induced constraint release and construct the new equations
which the segmental stretch and the tube survival probability
satisfy, respectively. We promote our contour-variable model of
constitutive equations in the end of this section. In section
\ref{sec. 3} we give the results and discussion, in which we focus
on the simple steady shear flow and the start-up of steady shear
flow. In section \ref{sec. 4}, we give the conclusions. The
process and methods used for the calculation of constitutive
equations are shown in the Appendix.

\section{The New Contour-Variable Model  \label {sec. 2}}

\subsection{Rate of CCR  \label {sec. 2.1}}
CCR represents the entanglements relaxation mechanics caused by
the convective motion of surrounding chains. It is not important
when the concentrated system is at rest or undergoes a slow flow.
However, when the velocity of the flow is comparable to the
inverse of reptation time of the primitive chain, CCR will play a
key role in determining the dynamics of the system. Taking into
account the CCR mechanics, Ianniruberto and Marrucci (1996)
proposed a contour location independent rate of CCR. In this
model, the rate of CCR is only caused by the retraction of the
primitive chain ends. So the rate of CCR is proportional to the
velocity gradient of the movement of the primitive chain ends.
However, not all convections or deformations will release
constraint. In order to exclude the case of affine deformations,
in which no constraint release occurs, Mead et al. gives the rate
of CCR as the difference between the rate of convection of the
entanglements and the rate of primitive chain retraction (Mead et
al., 1998; Viovy et al., 1983). They only evaluate the rate of CCR
at the chain ends, since they think the topological constraint can
be released only by the motion of chain ends (Ianniruberto and
Marrucci, 1996; Marrucci et al., 2001). (We call it the
end-relaxation mechanism in the follows.)

In the present work, we focus on the non-uniform behavior of CCR.
Let us consider a primitive chain segment confined by a tube
segment which was created at past time $t^{\prime}$, as called the
$t^{\prime}$-segment. The deformation of this chain segment will
be released when the entanglements forming the tube segments are
convected away by the flow. It is an irreversible process. This
$t^{\prime}$-segment will never be created again. When this
happens, some hidden entanglements become active (Ianniruberto and
Marrucci, 2000), which instantaneously makes the released part
(e.g. the $t^{\prime}$-segment) become a part of a new tube. Thus
we can say that the rate of creating a new segment is the same as
that of annihilating an old segment. The contribution to the
stress for a specified tube only comes from the part that remain
unreleased. The released part of the original tube, which becomes
a part of a new tube, will still have contributions to the stress.
However, the stress should be calculated from the new tube. Define
$s_{0}$ as the equilibrium contour variable, which runs from
$-L_{0}/2$ to $L_{0}/2$, where $L_{0}$ is the equilibrium length
of the primitive chain. When the chain is stretched, its primitive
length $L$ can be larger than $L_{0}$. We then define a local
stretch function $s(s_{0},t)$ which takes the value from $-L/2$ to
$L/2$, to describe the segmental stretch. We can define
$q(s_0,t)=\partial s/\partial s_{0}$ as a local strain. It
describes the local extension. The rate of CCR $k(s_{0},t)$ can be
written as the following form
\begin{equation}
 k(s_{0},t)=
 \mbox{\boldmath$\kappa$}:\mathbf{S}(s_{0},t)q(s_{0},t)-\frac{1}{q(s_{0},t)}
 \frac{\partial q(s_{0},t)}{\partial t},
\end{equation}
where \mbox{\boldmath$\kappa$} is the velocity gradient tensor of
the flow, which, we suppose, is homogeneous, although in general
it is not; $ \mathbf{S}(s_{0},t)$ is the orientational tensor,
$\mathbf{S}(s_{0},t)\equiv\langle\mathbf{u}(s_0,t)\mathbf{u}(s_0,t)
\rangle$. Apparently, it is a local approximation form. The first
term in Eq.~(1) describes the rate of entanglements convection at
contour variable $s_{0}$ and $t$. The second term is used to
exclude the case of affine deformation, in which there are no
environment changes with respect to the primitive chain and no CCR
happens. As $\mathbf{S}(\pm L_{0}/2,t)=0$, $q(\pm L_{0}/2,t)=1 $,
from Eq.~(1) we then find no CCR happens at the chain ends of the
primitive chain. In fact the ends of the tube change
instantaneously with the ends of the primitive chain during
movement, there is no relative motion between the ends of the tube
and the ends of the primitive chain. The primitive chain and the
tube are intrinsically synonymous, we treat them as different
things only for convenience in describing the motion of the chain
considered and the motion of its surrounding chains.

The difficulty of the flow-reversal problem has be pointed out
(Wapperom and Keunings, 2000; Ianniruberto and Marrucci, 2001). If
the present model reduces to the uniform and
single-relaxation-time version, this problem can be amended using
the absolute value of
$\left|\mbox{\boldmath$\kappa$}:\mathbf{S}\right|q$ in Eq.~(1). In
the present non-uniform case, we suppose that the above expression
can amend this difficulty. In the following we will not write the
absolute value sign in the rate of CCR by considering the flow
only in one direction.

\subsection{Segmental Stretch \label{sec. 2.2}}

The dynamics of segmental stretch is described by the equation of
motion, in which we neglect the acceleration term. It is derived
from the force balance between the segmental extension force
caused by the flow and the entropic elastic force. If the chain is
considered to be non-Gaussian, this equation can be written as
(Mead and Leal, 1995a, b)
\begin{equation}
\frac{\partial}{\partial t}s(s_{0},t) = \langle
v(s_0,t)\rangle+\frac{3\beta ZD}{\alpha}
\left[\frac{d}{dq}\mathcal{L}^{-1}(\alpha q)\right ]
\left(\frac{\partial^2 s}{\partial s_{0}^{2}}\right).
\end{equation}
Here $\langle v(s_0,t)\rangle$ is the relative pre-averaged
tangential velocity with respect to the center of chain
\begin{equation}
\langle v(s_0,t)\rangle=\mbox{\boldmath$\kappa$}:\int_0^{s_{0}}
\mathbf{S}(s^{\prime}_{0},t)q(s^{\prime}_{0},t)ds^{\prime}_0.
\end{equation}
The parameter $Z=M/M_e$ is the number of entanglements in a
primitive chain, where $M_e$ is the molecular weight between two
entanglements. $Z=T_d/3T_R$ (Doi and Edwards, 1986), where $T_d$
is the reptation time and $T_R$ is the Rouse relaxation time. $D$
defines the diffusion coefficient along the primitive path, and is
related to the reptation time by $T_d=L_0^2/\pi^2D $. $\beta$ is a
coefficient related to non-Gaussian behavior, which is given by
\begin{equation}
\beta=\frac{1-\alpha^2}{3-\alpha^2},
\end{equation}
where $\alpha=L_0/L_{max}$, and $L_{max}$ is the maximum length
the primitive chain can be stretched to. $\beta$ ensures that the
magnitude of a fictitious force acting along the primitive chain
is equal to $3k_BT/a$ when there is no segmental extension (Doi
and Edwards, 1986; Marrucci and Grizzuti, 1988). Here, $a$ is the
step length of the primitive chain and is a constant in the
present model. A good discussion for its magnitude is given by
Milner (2005). The inverse Langevin function $\mathcal{L}^{-1}(x)$
with the fractional extension $x=\alpha q$ is used to describe the
tension of a non-Gaussian finitely extensible polymer segment
(Treloar, 1975), which satisfies the following equation
\begin{equation}
\mathcal{L}(x)= \coth (x)- \frac{1}{x}.
\end{equation}
Expanding $\mathcal{L}^{-1}(x)$ in a Taylor series and using
Pad\'{e} approximation (Cohen, 1991), we can obtain
\begin{eqnarray}
\mathcal{L}^{-1}(x)&=&3x+\frac{9}{5}x^3+\frac{297}{175}x^5+\cdots \nonumber \\
 & \simeq &  x \frac{3-x^2}{1-x^2}.
\end{eqnarray}
 We differentiate both sides of Eq.~(2) with respect to $s_0$ and obtain
\begin{equation}
\frac{\partial q(s_0,t)}{\partial t}=
\mbox{\boldmath$\kappa$}:\mathbf{S}(s_0,t)q+\frac{3ZD\beta}{\alpha}
\frac{d\mathcal{L}^{-1}(\alpha q)}{dq}\frac{\partial^2 q}{\partial
s_{0}^{2}}+\frac{3ZD\beta}{\alpha} \left[\frac{d^2}{dq^2}
\mathcal{L}^{-1}(\alpha q)\right ]\left( \frac{\partial
q}{\partial s_0} \right )^2.
\end{equation}
Insert Eq.~(6) into the above equation, we obtain
\begin{equation}
\frac{\partial q}{\partial t}= \mbox{\boldmath$\kappa$}:\mathbf{S} q
+3ZD\beta \frac{3+\alpha^4q^4}{(1-\alpha^2q^2)^2} \frac{\partial^2
q}{\partial s_{0}^{2}}+3ZD\beta\alpha\frac{4\alpha^3q^3+12\alpha q}
{\left(1-\alpha^2q^2\right)^3}  \left( \frac{\partial q}{\partial
s_0} \right )^2.
\end{equation}
Taking into account the effect of CCR and following the argument
by Mead, Larson and Doi (1998), finally we obtain
\begin{eqnarray}
\frac{\partial q}{\partial t} &=& 3ZD\beta
\frac{3+\alpha^4q^4}{(1-\alpha^2q^2)} \frac{\partial^2q}{\partial
s_{0}^{2}}+3ZD\beta\alpha\frac{4\alpha^3q^3+12\alpha q}
{\left(1-\alpha^2q^2\right)^3}\left( \frac{\partial q}{\partial s_0}
\right )^2 \nonumber \\ & &+\mbox{\boldmath $\kappa$}: \mathbf{S}q -
\frac{1}{2}\left(\mbox{\boldmath$\kappa$}:
\mathbf{S}q-\frac{1}{q}\frac{\partial q}{\partial t}\right)(q-1).
\end{eqnarray}
The parameter $1/2$ comes from the fact that the magnitude of the
rate of entanglements reduction due to CCR is two times as large
as the rate of segmental retraction. The boundary condition and
the initial condition are
$$q(s_0,t)|_{s_0=\pm L_0/2}=1 $$ and $$q(s_0,t)|_{t=0}=1,$$ respectively.
The boundary condition is specified by the untethered condition of
the chain ends which can never be stretched.

\subsection{Probability of Tube Survival \label{sec. 2.3}}

The process of the segmental orientation can be described by the
tube survival probability $G(s,t,t^{\prime})$, which means that a
tube segment created at past time \emph{t}$^{\prime}$ still can be
found at location $s$ and time $t$. In the original DE model,
$G(s,t,t^{\prime})$ satisfies a diffusion equation. When the
segmental stretch is taken into account, such as the DEMG model, a
convective term has to be added to the DE model. This term
accounts for the flow induced convection of the \emph{defects} to
the ends of the primitive chain and the elongation of the tube
segments. However, the orientation relaxation induced by
constraint release has not been considered. Without that it makes
the prediction of the rheological properties by these models
qualitatively different from the experimental results.

Based on our understanding for the effect of constraint release,
we modify the CCR term in the equation of the probability of tube
survival in MLD model and drop the term which comes from the
reptative constraint release. Including the contribution from
chain stretch, the equation for the probability of the tube
survival is given by
\begin{equation}
\frac{\partial G(s,t,t^{\prime})}{\partial t}=D\frac{\partial^2
G(s,t,t^{\prime})}{\partial s^2}-\langle
v(s,t)\rangle\frac{\partial G(s,t,t^{\prime})}{\partial s},
\end{equation}
where $s=s(s_0,t)$ is the tube strain function. The boundary
condition and the initial condition are
 $$G(s,t,t^{\prime})|_{s=\pm L/2}=0$$ and $$ G(s,t,t^{\prime})|_{t=t^{\prime}}=1,
 $$respectively. It is convenient to transform the independent
variables $s$ and $t$ to other independent variables $s_0$ and $t$
(Pearson et al., 1991).
\begin{equation}
\frac{\partial G(s_0,t,t^{\prime})}{\partial t}=\frac{D}{q^2}
\frac{\partial^2 G(s_0,t,t^{\prime})}{\partial
s_0^2}+\left(-\frac{D}{q^3}\frac{\partial q}{\partial
s_0}-\frac{1}{q}\langle v(s_0,t)\rangle+\frac{1}{q}\frac{\partial
s}{\partial t}\right)\frac{\partial G(s_0,t,t^{\prime})}{\partial
s_0},
\end{equation}
where $q=q(s_0,t)$ is the function of local strain. When we consider
the CCR mechanism, a new term must be added to Eq.~(11), which is $$
\left(\mbox{\boldmath$\kappa$}: \mathbf{S}q-\frac{1}{q}
\frac{\partial q}{\partial t} \right)G(s_0,t,t^{\prime}). $$ Here,
$\mathbf{S}$ is the orientational tensor. Then we obtain
\begin{eqnarray}
\frac{\partial G(s_0,t,t^{\prime})}{\partial t}&=&\frac{D}{q^2}
\frac{\partial^2 G(s_0,t,t^{\prime})}{\partial
s_0^2}+\left(-\frac{D}{q^3}\frac{\partial q}{\partial
s_0}-\frac{1}{q}\langle v(s_0,t)\rangle+\frac{1}{q}\frac{\partial
s}{\partial t}\right)\frac{\partial G(s_0,t,t^{\prime})}{\partial
s_0} \nonumber \\ & &-f\left(\frac{\partial s}{\partial
s_0}\right)\left (\mbox{\boldmath$\kappa$}:
\mathbf{S}q-\frac{1}{q}\frac{\partial q}{\partial t}
\right)G(s_0,t,t^{\prime}),
\end{eqnarray}
where $ f (\partial s / \partial s_0)=1/q$ is the switch function
(Mead et al., 1998). When $q$ is large, the contribution from the
last term in the right hand side of Eq.~(12) is much smaller.
However, when $q$ approaches to unity, this term plays an
important role. This is arisen from the large difference in the
time scale between the reptation time and the Rouse retraction
time. It means that the relaxation of the tube orientation caused
by constraint release of surrounding chains should start to happen
just when the primitive chain nearly finish its retraction. The
boundary condition is
$$G(s_0,t,t^{\prime})|_{s_0=\pm L_0/2}=0, $$
and the initial condition is
$$ G(s_0,t,t^{\prime})|_{t=t^{\prime}}=1,$$
Note that $G(s_0,t,t^{\prime})$ is the probability function, which
has a range from 0 to 1.

\subsection{Constitutive Equations \label{sec. 2.4}}

The stress is the sum of tensile force of the primitive chain
projecting to some direction $\mathbf{u}$ by averaging over the
conformations of primitive chains. Those primitive chains
penetrate a unit area plane with its normal direction
$\mathbf{u}$.  The tensile force acting on segment $s_0$, along
the non-Gaussian primitive chain at time $t$ can be expressed by
the inverse Langevin function
\begin{equation}
F(s_0,t)=\frac{k_BT}{a}\mathcal{L}^{-1}\left(\alpha \frac{\partial
s}{\partial s_0} \right ).
\end{equation}
When the Pad\'{e} approximation is used, we have
\begin{equation}
F(s_0,t)=\frac{3k_BT}{a}\beta\frac{3-\alpha^2 q^2}{1-\alpha^2
q^2},
\end{equation}
where $\beta$ has the same meaning as in Eq.~(4), and is used to
conform that when $q=1$, $F(s_0,t)=3k_BT/a$ is equal to the
fictitious force acting on every segment (Marrucci and Grizzuti,
1988). The stress can be written as
\begin{eqnarray}
\mbox{\boldmath $\sigma$} &=& \frac{c}{N} \langle
\int^{L/2}_{-L/2}d s F(s_0,t) \mathbf{u}(s,t)\mathbf{u}(s,t)
\rangle \nonumber \\ & \simeq & \frac{c}{N}\int^{L/2}_{-L/2}d
s\langle F(s_0,t)\rangle \langle\mathbf{u}(s,t)\mathbf{u}(s,t)
\rangle \nonumber \\&=& \frac{15}{4}
\frac{G^0_N}{L_0}\int^{L_0/2}_{-L_0/2} \beta\frac{3-\alpha^2
q^2}{1-\alpha^2 q^2} q^2 \mathbf{S}(s_0,t) d s_0,
\end{eqnarray}
where $c$ is the number of polymers in a unit volume and \emph{N}
being the degree of polymerization. We decouple the average
segmental tension from the average segmental orientation in the
calculation of the first line to the second line. $G_N^0$ is the
plateau modulus, $G_N^0=4ck_{B}TL_{0}/5Na$. $\mathbf{S}(s_0,t)$
can be expressed by
\begin{equation}
\mathbf{S}(s_0,t)=\int^t_{-\infty} dt^{\prime} \frac{\partial
G(s_0,t,t^{\prime})}{\partial
t^{\prime}}\mathbf{Q}(\mathbf{E}(t,t^{\prime})).
\end{equation}
Here, $\mathbf{Q}$ is the DE strain tensor without the independent
alignment approximation given by
\begin{equation}
\mathbf{Q}=\left \langle \frac{\mathbf{\left(E\cdot u \right)
(E\cdot u) }} {|\mathbf{E\cdot u}|}\right\rangle_0
\frac{1}{\langle| \mathbf{E\cdot u}|\rangle_0},
\end{equation}
where $\mathbf{E}$ is the deformation tensor (Doi and Edwards,
1986).

\section{Results and discussion \label{sec. 3}}

The results reported in this paper will focus on steady and
start-up of simple shear flow of monodisperse concentrated
polymeric systems. We scale the equations for simplicity. The time
variable is scaled by the reptation time, $t/T_d \rightarrow t$,
and the spatial variable is scaled by the length of primitive
chain at equilibrium, $s_0/L_0 \rightarrow s_0$. For convenience,
we use the same symbol to denote the scaled quantities. Eqs.~(9),
(12) and (15) become
\begin{eqnarray}
\frac{\partial q}{\partial t} &=& \frac{3Z\beta}{\pi^2}
\frac{3+\alpha^4q^4}{(1-\alpha^2q^2)^2}
\frac{\partial^2q}{\partial
s_{0}^{2}}+\frac{3Z\beta\alpha}{\pi^2}\frac{4\alpha^3q^3+12\alpha
q} {\left(1-\alpha^2q^2\right)^3}\left( \frac{\partial q}{\partial
s_0} \right )^2 \nonumber \\
&&+\mbox{\boldmath$\kappa$}:\mathbf{S} q -
\frac{1}{2}\left(\mbox{\boldmath$\kappa$}:
\mathbf{S}q-\frac{1}{q}\frac{\partial q}{\partial t}\right)(q-1),
\end{eqnarray}
\begin{eqnarray}
\frac{\partial G(s_0,t,t^{\prime})}{\partial t}&=&\frac{1}{\pi^2
q^2} \frac{\partial^2 G(s_0,t,t^{\prime})}{\partial
s_0^2}+\left(-\frac{1}{\pi^2q^3}\frac{\partial q}{\partial
s_0}-\frac{1}{q}\langle v(s_0,t)\rangle+\frac{1}{q}\frac{\partial
s}{\partial t}\right)\frac{\partial G(s_0,t,t^{\prime})}{\partial
s_0} \nonumber \\ & &-\frac{1}{q}\left (\mbox{\boldmath $\kappa$}:
\mathbf{S}q-\frac{1}{q} \frac{\partial q}{\partial t}
\right)G(s_0,t,t^{\prime}),
\end{eqnarray}
\begin{equation}
 \mbox{\boldmath $\sigma$} =\int^{1/2}_{-1/2} \beta\frac{3-\alpha^2 q^2}
 {1-\alpha^2 q^2} q^2 \mathbf{S}(s_0,t) d s_0,
\end{equation}
respectively. Here, $\langle v(s_0,t)\rangle$ is given by Eq.~(3),
the relationship of which before scaling and after scaling is
$\langle v(s_0,t)\rangle T_d/L_0 \rightarrow \langle
v(s_0,t)\rangle$. The stress tensor has been scaled as \(
4\mbox{\boldmath $\sigma$}/15G^{0}_{N}\rightarrow \mbox{\boldmath
$\sigma$} \). $\mathbf{S}(s_0,t)$ is given by Eq.~(16). For the
case of shear flow,
$\mbox{\boldmath$\kappa$}=\dot{\gamma}\hat{e}_y\hat{e}_x$. The
shear viscosity is defined by $\eta=\sigma_{xy}/\dot{\gamma}$,
where $\sigma_{xy}$ is the shear stress and the primary normal
stress coefficient is given by
$\Psi=(\sigma_{xx}-\sigma_{yy})/\dot{\gamma}^{2}$, where
$\sigma_{xx}-\sigma_{yy}$ is the first normal stress difference.
The boundary conditions are
$$ q(s_0,t)|_{s_0=\pm 1/2}=1,$$
$$G(s_0,t,t^{\prime})|_{s_0=\pm 1/2}=0,$$
and the initial conditions are
$$ q(s_0,t)|_{t=0}=1,$$
$$G(s_0,t,t^{\prime})|_{t=t^{\prime}}=1, $$
These equations are nonlinear and have to be solved numerically.
Combining the finite differencing and the Newton-Raphson method,
we can obtain the solutions of these equations. The details are
shown in the Appendix. The constitutive parameters in these
equations include $Z$ and $\alpha$. Here, $Z$ is the number of
entanglements per chain and can be specified by fitting the steady
state predictions of the model and the experimental data. A larger
$Z$ corresponds to a shorter Rouse relaxation time. $\alpha$ is
the ratio between the equilibrium and the maximum length of the
primitive chain, reflecting the extensibility of the chain.

\subsection{Steady state shear flow \label{sec. 3.1}}

In this section, we show the dynamic properties of the chain
segments and rheological behavior under steady shear flow.

Fig.~1 gives the non-uniform segmental stretch of the primitive
chain with different shear rates. The parameters are chosen as
$Z=20$ and $\alpha=0$ which denote the Gaussian chain. Due to the
untethered fact of the chain ends, the segments at chain ends
remain unstretched. There exists a maximum value of stretch at the
center. The extension of the segments equivalently reduces the
diffusion coefficient by $1/q^{2}$ in Eq.~(19). So the equivalent
reduction of diffusion is much more at the center. In addition to
a small value of the rate of segmental convection around the
center, we find that the rate of tube relaxation, or the rate of
tube creation, is relatively small at the center. The cancellation
of gradually increasing rate of convection and decreasing rate of
effective diffusion from the center to ends induces a flat region
of the rate of the tube relaxation around the center. According to
the calculation from Eq.~(16) under steady state shear, we find
that around the center of the tube the value of $S_{xy}$ is
relatively small. At the ends, where the segments of tube
disappear at their creating time, $S_{xy}(\pm 1/2,t)=0$.
Therefore, there must be a maximum value of $S_{xy}$ in some
location between the center and the ends. This can be seen from
Fig.~2 when the shear rate is high enough. It is very interesting
that when the shear rate is high enough, the maximum contribution
to stress from segmental orientation comes neither from the ends
of the chain nor from the center, but from some other location in
between. Moreover, this location will move towards the ends when
the shear rate grows higher. From the calculation we also find
that the segmental orientation is insensitive to the entanglement
number of a primitive chain, denoted as $Z$ which determines the
segmental extension in steady state. Fig.~3 shows the differences
of $S_{xy}$ between the present model and DEMG model for $Z=20$
and $Z=50$, respectively. The result shows that including of the
CCR term will obviously enhance the value of $S_{xy}$. Therefore,
the magnitude of the stress in fast flow is seriously
underestimated in the DEMG model.

The comparisons of the shear stress and the steady state shear
viscosity with different values of $Z$ are given in Fig.~4. In the
linear region, where the steady shear rate is smaller than the
inverse of reptation time, both the shear stresses and steady
state shear viscosity are independent of $Z$, because in these
cases the shear rates are not high enough to stretch the chain
segments. When the shear rate is high enough, the segments will be
stretched to many times of their equilibrium length. The larger
extension of the segment is, the more it contributes to the shear
stress. On the other hand, the Rouse relaxation time $T_R$
determines the magnitude of segmental extension. If $T_R$ is
larger, which corresponds to a smaller $Z$, the shear stress will
be larger. Fig.~4 also shows the shear viscosity of the steady
state which is defined as
$\eta(\dot{\gamma})=\sigma_{xy}/\dot{\gamma}$. When the shear rate
is smaller than the inverse of reptation time, it behaves as
Newtonian fluid. While in the high-shear-rate region it markedly
depends on the shear rate. Different segmental stretch results in
the differences of the first normal stress difference in the
high-shear-rate region, which can be seen in Fig.~5. The same
characteristics for the primary normal stress coefficients are
also shown in Fig.~5.

Up to now the chains are approximated as Gaussian chains
($\alpha=0$), which is valid if the stains are not too large.
However, under fast flow the chains are largely stretched and
approach to their limiting extensible values, the Gaussian
treatment is no longer valid. In the following, we will use the
non-Gaussian treatment of the single chain taking into account the
finite extensibility of the chain ($\alpha\neq0 $). Fig.~6 shows
the comparisons of the shear stress and the first normal stress
difference with respect to the shear rate with different
extensibilities for $\alpha=0,0.2,0.4$ and $0.6$, respectively.
The parameter $Z$ is chosen as $Z=20$. The calculation also
indicates that for a specified value of the shear rate segmental
orientation is not sensitive to segmental extensibility. From
Fig.~6 it can be found that the curves of shear stress and first
normal stress difference are independent of segmental
extensibility in the small shear rate region where the stress
contribution from segmental stretch can be ignored. In the
high-shear-rate region, the differences of shear stress and the
differences of the first normal stress caused by the differences
of segmental extensibility are mainly contributed from segmental
extension. The segmental orientation is only determined by the
value of the shear rate, and is independent of the segmental
stretch, because they are considered as independent processes.

Fig.~6 also shows the comparisons between the predictions of the
present model and the experimental data (Pattamaprom and Larson,
2001) of the stresses and the first normal stress differences
under steady shear for different parameters of $\alpha$; Fig.~7
shows the comparisons with experimental data of the stresses and
the first normal stress differences under steady shear for
different parameters of $Z$. By fitting the experimental data, we
find that incorporating the non-uniform features to the
constitutive models does not affect rheological properties.
Moreover, the non-uniform model gives us the detailed descriptions
of the chain dynamics. In the present model there are two
modifiable parameters of $Z$ and $\alpha$. By fitting the
experimental data to the predictions of the present model, we can
obtain the values of $Z$ and $\alpha$, respectively. The
comparison between them will show you the relative abilities of
the retraction and the extension of the chains. This means that,
under the same magnitude of shear rate, different abilities of
retraction or extension will result in different magnitudes of
stresses, although the molecules composing the materials may have
the same volume fractions and molecular weights. From these two
figures, quantitative deviation from the experimental data exists,
especially the first normal stress difference in the
high-shear-rate region. These deviations may attribute to the
``nonlocal" interactions between chains. By comparing with the
uniform models by Pattamaprom and Larson (2001), we also find that
the effect of non-uniform CCR on the stress and the normal stress
is minor. The details of segmental orientation and stretch are not
important in determining the rheological properties in steady
simple shear.

\subsection{Start-up of Steady Shear Flow \label{sec. 3.2}}

The transient behavior of chain stretch and orientation, as well
as the rheological properties, are discussed in this section. The
steady shear is imposed to the system at $t=0$.

In Fig.~8 the segmental stretch varying with time at different shear
rates are shown, and the constitutive parameters are chosen as
$Z=20,\alpha=0.0 $. The contour variable is focused at $s_0=0$.
Before this segment reaches its steady value of extension, it passes
a maximum value which is dependent on the shear rate, since other
parameters are fixed. This maximum value will appear earlier when
the shear rate increases. Fig.~9 gives the behavior of segmental
orientation $S_{xy}(0,t)$ with the constitutive parameters chosen as
the same in Fig.~8. We focus on the location $s_0=0$. If the shear
rate is of the order of $1/T_d$ or larger, the value of
$S_{xy}(0,t)$ will also pass a maximum value before reaching its
steady value. The appearance of the maximum value is not due to the
fact of segmental stretch, but the existence of a maximum value in
the strain $Q_{xy}$ with respect to the time.

Fig.~10 and Fig.~11 show the transient behavior of the shear stress
and the first normal stress difference, respectively, where
overshoots are predicted in the present model. For the shear stress,
when the shear rate is higher than the inverse of reptation time,
the overshoot appears. The overshoot of the first normal stress
difference appears at a much higher shear rate in the present model
than that in other models. The reason is that the appearance of
maximum in the first normal stress difference is determined by the
evolvement of segmental stretch which should also have a maximum in
the course of time. However, the rate of CCR we obtained contains a
local strain $q(s,t)$, which largely decreases the maximum value of
segmental stretch and subsequently the stress. Thus the magnitude of
the start-up shear rate, which can produce a overshoot, increases.

Fig.~12 and Fig.~13 show the comparisons of the shear viscosities
and the first normal stress differences with the experimental
results, respectively. We choose the shear viscosities instead of
the shear stresses for convenience to compare with the
experimental data (Pattamaprom and Larson, 2001), in which the
material is sample one. The parameters in these two figures are
both chosen as $Z=20$ and $\alpha=0$, $0.3$, $0.6$, respectively.
The results indicate that only when the shear rate is high, i.e.
$\dot {\gamma}\geq 10/T_d$, the differences caused by the
different values of $\alpha$ are predicted by the present model.
If the chains are less extensible, which corresponds to a large
value of $\alpha$, the segmental extension will be smaller under
the same shear rate at the same time. This leads to smaller shear
viscosities as well as the first normal stress differences. In
fact, the segmental orientation is mainly determined by the
magnitude of the start-up shear rate. When the shear rate is fixed
the contribution from orientation to shear stress is specified.
Thus, the magnitude of segmental stretch plays a critical role in
determining the final value of the shear stress, and subsequently
the shear viscosity. This is kept when the first normal stress
difference is considered. By fitting the experimental data we find
that the predictions of the present model have the same tendency
with the experimental curves. On the other hand, there are large
inconsistencies when the quantitative values are concerned,
especially the magnitude of overshoots. Both the shear stress and
the normal stress overshoots in the present model are very weak
comparing to the experimental data. It is obvious that the chain
stretch is not enough, which may arise from the inclusion of local
approximation in CCR in the present model.

\section{Conclusions \label{sec. 4}}

In this paper, we develop a contour-variable model to describe the
non-uniform behavior of segments under steady shear and start-up
of steady shear, and the subsequent contributions to the shear
stress and the first normal stress difference. The present model
is a modification of the MLD model with a modified CCR term and
non-Gaussian chain treatment of the single chain. We drop the
contour length fluctuation contribution and neglect the reptative
constraint release term, since both of them contribute less than
the shear rate dependent CCR term in the high-shear-rate region.
The numerical results predict strong dependencies of the segmental
stretch and orientation on the contour variable. The maximum
extension happens in the center of the chain due to the demand of
symmetry. The fact that segmental extension passes a maximum
during the start-up of steady shear flow is the key reason why the
first normal stress difference has a maximum. The segmental
orientation also shows a maximum in the center when shear rate is
small, and this is mainly attributed to the reptation. However,
when the shear rate is high enough the segmental convection begins
to work. A maximum of segmental orientation appears in some
locations between the center and the end, where the stress
contribution from orientation is the largest. The inclusion of CCR
entirely promotes the magnitude of orientation of each segment
rather than changes the distribution of the value of segmental
orientation. The non-uniform segmental orientation and stretch
subsequently contribute to rheological properties, which
qualitatively agree with experimental data, although quantitative
deviation still exists. When the shear rate is high, i.e.
$\dot{\gamma}\gg T_d^{-1}$, the segmental strain is larger than 1.
The difference of the segmental extensibilities is the main reason
for the difference of rheological behavior. The same properties
hold in the start-up of steady shear. This validates our
pre-approximation of the independence of segmental stretch and
orientation.

A few approximations have been used in the present model. One is
the local approximation of CCR. CCR happens once a tube segment is
convected away from its original location, since the relative
position to the primitive-chain segment confined by this tube
segment changes. The convection of a segment may not be only
determined by its local conditions, but all other segments
elsewhere. In fact, the present model only predicts the overshoots
in shear viscosity and first normal stress difference under the
start-up of steady shear, but does not predict undershoots which
have been observed in experiments (Huppler et al., 1967; Mhetar
and Archer, 2000). It is well known that the appearance of
overshoots is due to segmental stretch. How about the undershoots?
Pattamaprom and Larson (2001) predicted that the undershoots would
appear if the dependence of the orientation and stretch on the
contour variable was taken into account. The present model
indicates that the inclusion of ``local'' non-uniform rate of CCR
to the constitutive equations does not predict undershoots.
Undershoots may appear if we include the nonlocal interaction into
the model. That is the coupling between chains, e.g. the
hydrodynamic interactions, which are neglected in the present
model. Another is that we suppose the shear flow is homogeneous,
which in general is not. At last we can say that in polymer melts
under shear non-uniform segmental orientation and stretch are hard
to be observed directly in experiments, but their effect on
rheological behavior can be detected, though the influences of the
microscopic non-uniform behavior on them is not remarkable.
However, the processes of crystallization and the dynamics of
phase separation (Maurits and Fraaije, 1997) should be
significantly affected by these contour dependent quantities, as
the segmental dynamics plays a crucial rule in these processes. In
fact the kinetic coefficients are related to the non-uniform
features (Kawasaki and Sekimoto, 1988; Kawakatsu, 1997). It is of
great significance to investigate the segmental dynamics and their
affiliated properties theoretically by microscopic or mesoscopic
models.

\bigskip
\begin{center}
\textbf{ACKNOWLEDGMENTS}
\end{center}

We thank Dr.~Bing Miao for the helpful discussion and suggestions.
This work is supported by National Natural Science Foundation of
China (NSFC) 20490222, 20874111, 50821062 and the Grant from
Chinese Academy of Sciences KJCX2-YW-206.

\newpage
\appendix
\section{}

In this Appendix we describe the process and methods for the
calculation of constitutive equations. The DE strain tensor
without the independent alignment approximation, i.e. Eq.~(17), is
integrated numerically. The unit vector of orientation is chosen
as $\mathbf{u}=(\sin\theta\cos\varphi, \sin\theta\sin\varphi,
\cos\theta)$
\begin{eqnarray}
Q_{xy} &=& \left \langle \frac{(\mathbf{E\cdot u})_{x}
(\mathbf{E\cdot u})_{y} } {|\mathbf{E\cdot u}|}\right\rangle_0
\frac{1}{\langle| \mathbf{E\cdot u}|\rangle_0} \nonumber \\ &=&
\left \langle\frac{(u_x+\gamma u_y)\cdot u_y}{(1+2\gamma
u_xu_y+\gamma^2 u_y^2)^{1/2}} \right
\rangle_0\frac{1}{\langle(1+2\gamma u_xu_y+\gamma^2
u_y^2)^{1/2}\rangle_0},
\end{eqnarray}
where $\gamma$ is the shear deformation, $\left<\ldots\right>_{0}$
denotes the average over the isotropic state, i.e.,
$\left<\ldots\right>_{0}=\int d\mathbf{u}/4\pi(\ldots). $ Then we
have
\begin{eqnarray}
\alpha(\gamma)&\equiv&\langle(1+2\gamma u_xu_y+\gamma^2
u_y^2)^{1/2}\rangle_0 \nonumber \\ &=& \frac{1}{4\pi}\int^{\pi}_0
\sin\theta d\theta\int^{2\pi}_0
d\varphi(1+2\gamma\sin^2\theta\sin\varphi\cos\varphi+
\gamma^2\sin^2\theta\sin^2\varphi)^{1/2} \nonumber \\ &=&
\int^1_0dx\int^1_0dy
\left[1+\frac{1}{2}\gamma^2(1-x^2)+\gamma(1-x^2)\sin(4\pi y)
-\frac{1}{2}\gamma^2(1-x^2)\cos(4\pi y) \right ]^{1/2}.
\end{eqnarray}
Here, the variable transformations, i.e. $x=\cos\theta$,
$y=\varphi/2\pi$, are introduced. Then we have
\begin{equation}
Q_{xy}=\frac{1}{\alpha(\gamma)}\frac{1}{2}\int_0^1dx\int_0^1dy
\frac{(1-x^2)\sin(4\pi y)+\gamma(1-x^2)-\gamma(1-x^2)\cos(4\pi
y)}{\left[1+\frac{1}{2}\gamma^2(1-x^2)+\gamma(1-x^2)\sin(4\pi y)
-\frac{1}{2}\gamma^2(1-x^2)\cos(4\pi y) \right ]^{1/2}}.
\end{equation}
With the same procedure, we obtain
\begin{eqnarray}
Q_{xx}-Q_{yy} &=& \frac{1}{\alpha(\gamma)}\int_0^1dx\int_0^1dy
\nonumber \\ &\times& \frac{\left(1-\frac{1}{2}\gamma^2
\right)(1-x^2)\cos(4\pi y)+\gamma(1-x^2)\sin(4\pi
y)+\frac{1}{2}\gamma^2(1-x^2)}{\left[1+\frac{1}{2}\gamma^2(1-x^2)+
\gamma(1-x^2)\sin(4\pi y) -\frac{1}{2}\gamma^2(1-x^2)\cos(4\pi y)
\right ]^{1/2}}.
\end{eqnarray}
The two-dimensional integrals of Eqs. (A2)-(A4) are integrated out
using the extended \emph{Simpson's rule} in each dimension,
respectively.

The constitutive equations under start-up of steady shear are
solved numerically. We discretize Eq.~(18) using the
Crank-Nicolson scheme, and define $F$ as
\begin{eqnarray}
F=u(i,j+1)-q(i,j+1)+u(i,j)+q(i,j),
\end{eqnarray}where \emph{i} and \emph{j} denote the discretized contour
variable grid points and time variable grid points, respectively.
Here, $u(i,j)$ is given by
\begin{eqnarray}
u(i,j) &=&\frac{3Z\beta}{\pi^2}\frac{\Delta t}{(\Delta s)^2}
\frac{3+\alpha^4q^4(i,j)}{[1-\alpha^2q^2(i,j)]^2}\frac{q(i,j)}{1+q(i,j)}
\left[q(i+1,j)-2q(i,j)+q(i-1,j)\right] \nonumber \\ & &+
\frac{3Z\beta\alpha}{\pi^2} \frac{\Delta t}{(\Delta s)^2}
\frac{4\alpha^3q^3(i,j)+12\alpha q}{\left[1-\alpha^2
q^2(i,j)\right]^3}\frac{q(i,j)}{1+q(i,j)}
\frac{\left[q(i+1,j)-q(i-1,j)\right]^2}{4} \nonumber \\ & &
+\frac{1}{2}\Delta t\dot \gamma S_{xy}(i,j)
\frac{3q^2(i,j)-q^3(i,j)}{1+q(i,j)}
\end{eqnarray}
where $\Delta s$ and $\Delta t$ are the contour and time grid
steps, respectively. The contour grid step is a constant in our
calculation, while there are two kinds of time step if the shear
rate is higher than $10/T_d$. The grids density around the maximum
value of $Q_{xy}(\gamma)$ is larger than that the place far from
it, so that the effects arising from the sharp changes of the
value of $Q_{xy}(\gamma)$ around the maximum value can be
observed. There are $N_s-1$ coupled equations if we set $F=0$ for
each time grid, and $N_s$ is the grid number of of the contour
coordinate. The Newton-Raphson method is used to find their roots
at each time grid. The number of time step is $N_t$. In this
paper, we choose $\Delta s=5\times10^{-3}$ and $\Delta t$ has a
range from $5\times10^{-6}$ to $4\times10^{-4}$ depending on the
shear rate. For a larger shear rate, we choose a smaller $\Delta
t$. $N_t$ is determined by the time grid steps. Eq.~(19) can be
written as
\begin{equation}
\frac{\partial G(s_0,t)}{\partial t}=A_0(s_0,t)\frac{\partial^2
G(s_0,t)}{\partial s_0^2}-A_1(s_0,t)\frac{\partial
G(s_0,t)}{\partial s_0}+A_3(s_0,t)G(s_0,t),
\end{equation}
where
$$A_0(s_0,t)=1/(\pi^2q^2),$$ $$A_1(s_0,t)=\left(\frac{1}{\pi^2q^3}\frac{\partial
q}{\partial s_0}+\frac{1}{q}\langle
v(s_0,t)\rangle-\frac{1}{q}\frac{\partial s}{\partial t}\right),$$
$$A_3(s_0,t)=-\frac{1}{q}\left
(\mbox{\boldmath$\kappa$}:\mathbf{S}q-\frac{1}{q} \frac{\partial
q}{\partial t} \right).$$ Eq.~(A7) can be discretized as
\begin{eqnarray}
\frac{G(i,j+1)-G(i,j)}{\Delta t} &=& A_0(i,j+1)\frac{1}{2} \left[
\left(\frac {\partial^2 G} {\partial
s_0^2}\right)_{(i,j)}+\left(\frac{\partial ^2 G}{\partial
s_0^2}\right)_{(i,j+1)} \right ] \nonumber \\ & &
-A_1(i,j+1)\frac{1}{2} \left[\left(\frac{\partial G}{\partial
s_0}\right)_{(i,j)} +\left(\frac{\partial G}{\partial
s_0}\right)_{(i,j+1)} \right] \nonumber \\ & &
+A_3(i,j+1)\frac{1}{2} \left(G(i,j)+G(i,j+1)\right),
\end{eqnarray}
where
$$\left(\frac{\partial^2 G} {\partial
s_0^2}\right)_{(i,j)}=\frac {G(i+1,j)-2G(i,j)+G(i-1,j)}{(\Delta
s)^2},$$
$$\left(\frac{\partial G} {\partial
s_0}\right)_{(i,j)}=\frac{G(i+1,j)-G(i-1,j)}{2\Delta s}.$$

The complete procedure is as follows. For every time step, we
first assume that $S_{xy}(i,j+1)=S_{xy}(i,j)$. We solve the
nonlinear equations of $F=0$ ($F$ is given by Eq.~(A5)), then we
obtain $q(i,j+1)$. Insert these into Eq.~(A8) we obtain
$G(i,j+1)$. From Eq.~(16) we obtain the new values of
$S_{xy}(i,j+1)$. By inserting these values back to Eq.~(A5) and
let $F=0$, we obtain the new values of $q(i,j+1)$, then new
$G(i,j+1)$, and subsequently $S_{xy}(i,j+1)$. Repeat the above
steps until the expected convergent value ($10^{-12}$) is reached.
The calculation of constitutive equations under steady shear is
much easier than that under the start-up of steady shear, as in
the former case the orientational tensor $\mathbf Q$ and strain
$q$ are time independent, and the calculation procedure are the
same as in the case of start-up of steady shear at a certain time
grid.

\newpage


\textbf{References}

\begin{verse}
\setlength{\parskip}{0.ex} \setlength{\baselineskip}{14.5pt}

Cohen A (1991) A Pad\'{e} approximant to the inverse Langevin
function. Rheol Acta 30: 270-273

de Gennes PG (1971) Reptation of a polymer chain in the presence
of fixed obstacles. J Chem Phys 55: 572-579

de Gennes PG (1979) Scalling concepts in polymer physics. Cornell
University Press

des Cloizeaux J (1988) Double reptation vs. simple reptation in
polymer melts. Europhys Lett 5: 437-442

des Cloizeaux J (1990) Relaxation of entangled polymers in melts.
Macromolecules 23: 3992-4006

Doi M (1980) A constitutive equation derived from the model of Doi
and Edwards for cocentrated polymer solutions and polymer melts. J
Polym Sci Polym Phys Ed 18: 2055-2067

Doi M, Edwards SF (1978a) Dynamics of concentrated polymer
systems. Part 1.---Brownian motion in the equilibrium state. J
Chem Soc Faraday Trans II 74: 1789-1801

Doi M, Edwards SF (1978b) Dynamics of concentrated polymer
systems. Part 2.---Molecular motion under flow. J Chem Soc Faraday
Trans II 74: 1802-1817

Doi M, Edwards SF (1978c) Dynamics of concentrated polymer
systems. Part 3.---The constitutive equation. J Chem Soc Faraday
Trans II 74: 1818-1832

Doi M, Edwards SF (1979) Dynamics of concentrated polymer systems.
Part 4.---Rheological properties. J Chem Soc Faraday Trans II 75:
38-54

Doi M, Edwards SF (1986) The theory of polymer dynamics. Oxford
Science Publications

Graham RS, Likhtman AE, Mcleish TCB, Milner ST (2003) Microscopic
theory of linear, entangled polymer chains under rapid deformation
including chain strecth and convective constraint release. J Rheol
47: 1171-1200

Huppler JD, Macdonald IF, Ashare E, Spriggs TW, Bird RB (1967)
Rheological properties of three solutions. Part II. relaxation and
growth of shear and normal stresses. Trans Soc Rheol 11: 181-204

Ianniruberto G, Marrucci G (1996) On compatibility of the Cox-Merz
rule with the model of Doi and Edwards. J Non-Newtonian Fluid Mech
65: 241-246

Ianniruberto G, Marrucci G (2000) Convective orientational renewal
in entangled polymers. J Non-Newtonian Fluid Mech 95: 363-374

Ianniruberto G, Marrucci G (2001) A simple constitutive equation
for entangled polymers with chain stretch. J Rheol 45: 1305-1318

Janeschitz-Kriegl H (1982) Polymer melt rheology and flow
birefringence. Springer-Verlag, Berlin

Kawakatsu T (1997) Effects of changes in the chain conformation on
the kinetics of order-disorder transitions in block copolymer
melts. Phys Rev E 56: 3240-3250

Kawasaki K, Sekimoto K (1988) Morphology dynamics of block
copolymer systems. Physica 143A: 361-413

Larson RG (1984) A constitutive equation for polymer melts based
on partially extending strand convection. J Rheol 28: 545-571

Larson RG (1988) Constitutive equations for polymer melts and
solutions. Butterworths. Guildford

Leygue A, Bailly C, Keunings R (2005) A differential formulation
of thermal constraint release for entangled linear polymers. J
Non-Newtonian Fluid Mech 128: 23-28

Leygue A, Bailly C, Keunings R (2006a) A differential tube-based
model for predicting the linear viscoelastic moduli of
polydisperse entangled linear polymers. J Non-Newtonian Fluid Mech
133: 28-34

Leygue A, Bailly C, Keunings R (2006b) A tube-based constitutive
equation for polydisperse entangled linear polymers. J
Non-Newtonian Fluid Mech 136: 1-16

Marrucci G (1985) Relaxation by reptation and tube enlargement: a
model for polydisperse polymers. J Polym Sci Polym Phys Ed. 23:
159-177

Marrucci G (1986) The Doi-Edwards model without independent
alignment. J Non-Newtonian Fluid Mech 21: 329-336

Marrucci G (1996) Dynamics of entanglements a nonlinear model
consistent with the Cox-Merz rule. J Non-Newtonian Fluid Mech 62:
279-289

Marrucci G, Greco F, Ianniruberto G (2001) Integral and
differential constitutive equations for entangled polymers with
simple versions of CCR and force balance on entanglements. Rheol
Acta 40: 98-103

Marrucci G, Grizzuti N (1988) Fast flows of concentrated polymers
predictions of the tube model on chain stretching. Gazz Chim Ital
118: 179-185

Maurits NM, Fraaije JGEM (1997) Mesoscopic dynamics of copolymer
melts: from density dynamics to external potential dynamics using
nolocal kinetic coupling. J Chem Phys 107: 5879-5889

Mead DW, Larson RG, Doi M (1998) A molecular theory for fast flows
of entangled polymers. Macromolecules 31: 7895-7914

Mead DW, Leal LG (1995a) The reptation model with segmental
stretch I. basic equations and general properties. Rheol Acta 34:
339-359

Mead DW, Leal LG (1995b) The reptation model with segmental
stretch II. steady flow properties. Rheol Acta 34: 360-383

Mhetar VR, Archer LA (2000) A new proposal for polymer dynamics in
steady shearing flows. J Polym Sci Part B Polym Phys 38: 222-233

Milner ST (2005) Predicting the tube diameter in melts and
solutions. Macromolecules 38: 4929-4939

Milner ST, Mcleish TCB, Likhtman AE (2001) Microscopic theory of
convective constraint release. J Rheol 45: 539-563

Pattamaprom C, Larson RG (2001) Constraint release effects in
monodisperse and bidisperse polystyrenes in fast transient
shearing flows. Macromolecules 34: 5229-5237

Pearson DS, Helfand E (1984) Viscoelastic properties of
star-shaped polymers. Macromolecules 17: 888-895

Pearson DS, Herbolzheimer EA, Grizzuti N, Marrucci G (1991)
Transient behavior of entangled polymers at high shear rates. J
Poly Sci Phys Ed 29: 1589-1597

Pearson DS, Kiss AD, Fetters LJ, Doi M (1989) Flow-induced
birefringence of concentrated polyisoprene solutions. J Rheol 33:
517-535

Treloar LRG (1975) The physics of rubber elasticity, 3rd edn.
Clarendon Press. Oxford

Viovy JL, Monnerie L, Tassin JF (1983) Tube relaxation: a
necessary concept in the dynamics of strained polymers. J Polym
Sci Polym Phys Ed 21: 2427-2444

Wapperom P, Keunings R (2000) Simulation of linear polymer melts
in transient complex flow. J Non-Newtonian Fluid Mech 95: 67-83

Watanabe H (1999) Viscosity and dynamics of entangled polymers.
Prog Polym Sci 24: 1253-1403

\end{verse}

\newpage
{\bf Figure Captions}

\par
{\bf Fig.~1}  Non-uniform segmental stretch with different shear
rates under steady shear. $\dot\gamma$=0.1, 1, 10, 100, 500,
respectively, and $Z=20$, $\alpha=0$.

\par
{\bf Fig.~2}  Non-uniform segmental orientation with different
shear rates under steady shear. $\dot\gamma$=0.001, 0.1, 1, 10,
100, 500, respectively, and $Z=20$, $\alpha=0$.

\par
{\bf Fig.~3}  Comparison of non-uniform segmental orientation of
DEMG model and the present model under steady shear for $\alpha=0$
and $Z$=20, 50, respectively.

\par
{\bf Fig.~4}  Steady shear viscosity and shear stress for
$\alpha=0$ and $Z$=10, 20, 30, 50, respectively.

\par
{\bf Fig.~5}  First normal stress differences and primary normal
stress coefficients under steady shear for $\alpha=0$ and $Z$=10,
20, 30, 50, respectively .

\par
{\bf Fig.~6}  Comparison of the stresses and the first normal
stress differences under steady shear between the present model
(lines) and the experimental data (closed symbols). The parameters
are chosen as $Z=20$ and $\alpha=0$, $0.2$, $0.4$, $0.6$,
respectively.

\par
{\bf Fig.~7}  Comparison of the stresses and the first normal
stress differences under steady shear between the present model
(lines) and the experimental data (closed symbols). The parameters
are chosen as $\alpha=0$ and $Z=10$, $20$, $30$, $50$,
respectively.

\par
{\bf Fig.~8}  The transient behavior of segmental stretch at $s_0=0$
for $\dot\gamma$=0.1, 1, 10, 50, 100, respectively, and $Z=20$,
$\alpha=0$.

\par
{\bf Fig.~9}  The transient behavior of segmental orientation at
$s_0=0$ for $\dot\gamma$=0.01, 0.1, 1, 10, 100, 200, respectively,
and $Z=20$, $\alpha=0$.

\par
{\bf Fig.~10}  The transient behavior of shear stress under
different shear rates for $\dot\gamma$=0.001, 0.01, 0.1, 1, 10, 100,
respectively, and $Z=20$, $\alpha=0$.

\par
{\bf Fig.~11}  The transient behavior of the first normal stress
differences under different shear rates for $\dot\gamma$=0.1, 1, 10,
100, 200, respectively, and $Z=20$, $\alpha=0$.

\par
{\bf Fig.~12}  Comparison of the shear viscosities with the
experimental data (closed symbols). The parameters are chosen as
$Z=20$ and $\alpha=0$, $0.3$, $0.6$, respectively.

\par
{\bf Fig.~13}  Comparison of the first normal stress differences
with the experimental data (closed symbols). The parameters are
chosen as the same as those in Fig.~12.

\newpage

\clearpage

\begin{figure}
\centerline{\includegraphics[angle=0,scale=1,draft=false]{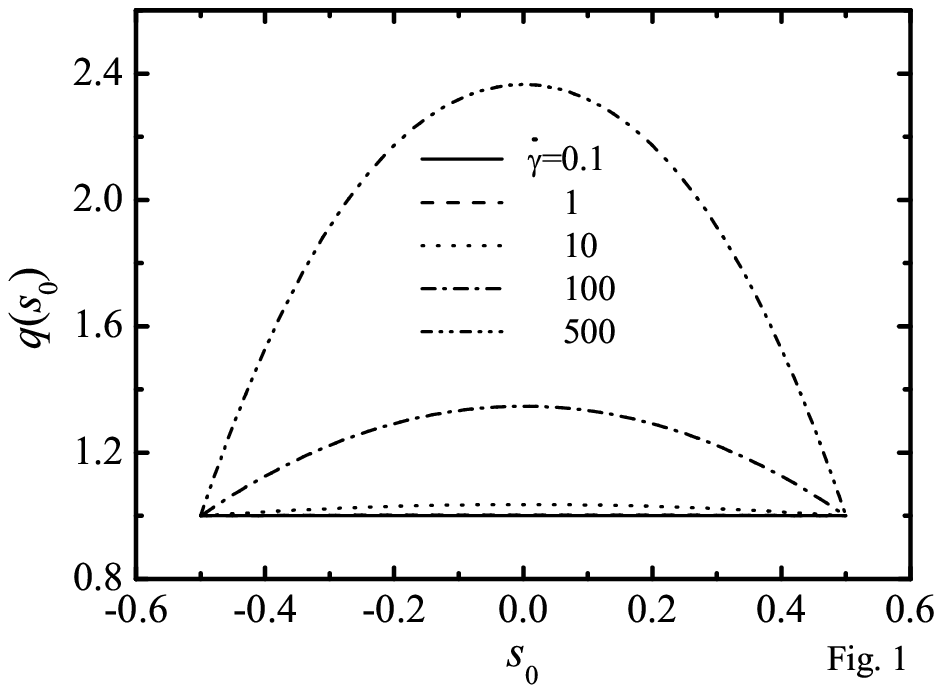}}
\label{01}
\end{figure}

\clearpage
\begin{figure}
\centerline{\includegraphics[angle=0,scale=1,draft=false]{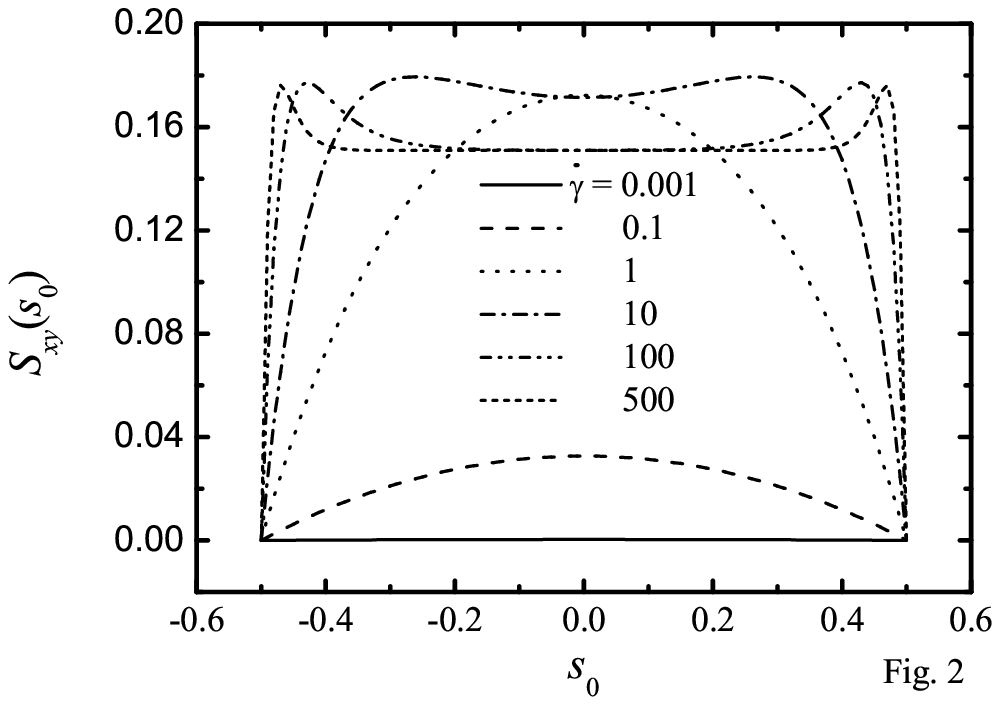}}
\label{02}
\end{figure}

\clearpage
\begin{figure}
\centerline{\includegraphics[angle=0,scale=1,draft=false]{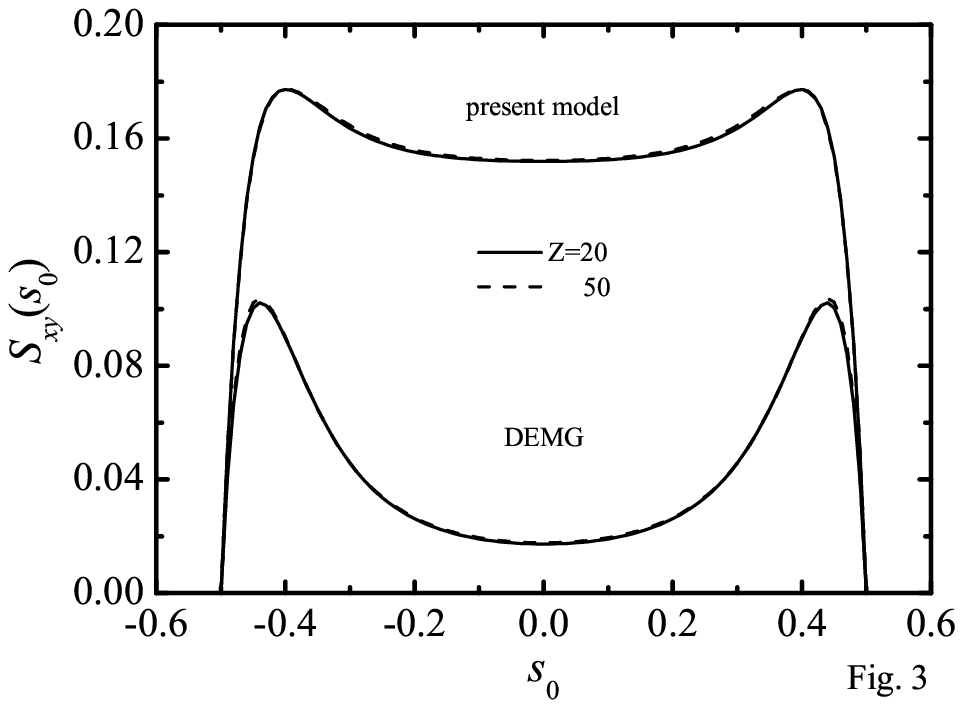}}
\label{03}
\end{figure}

\clearpage
\begin{figure}
\centerline{\includegraphics[angle=0,scale=1,draft=false]{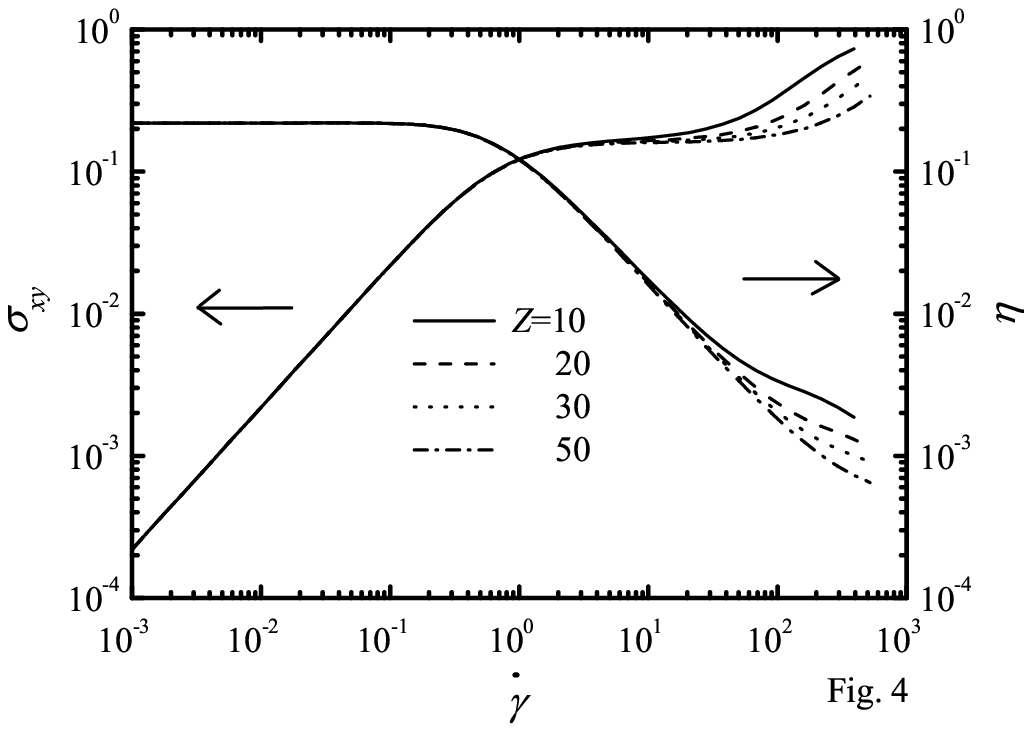}}
\label{04}
\end{figure}

\clearpage
\begin{figure}
\centerline{\includegraphics[angle=0,scale=1,draft=false]{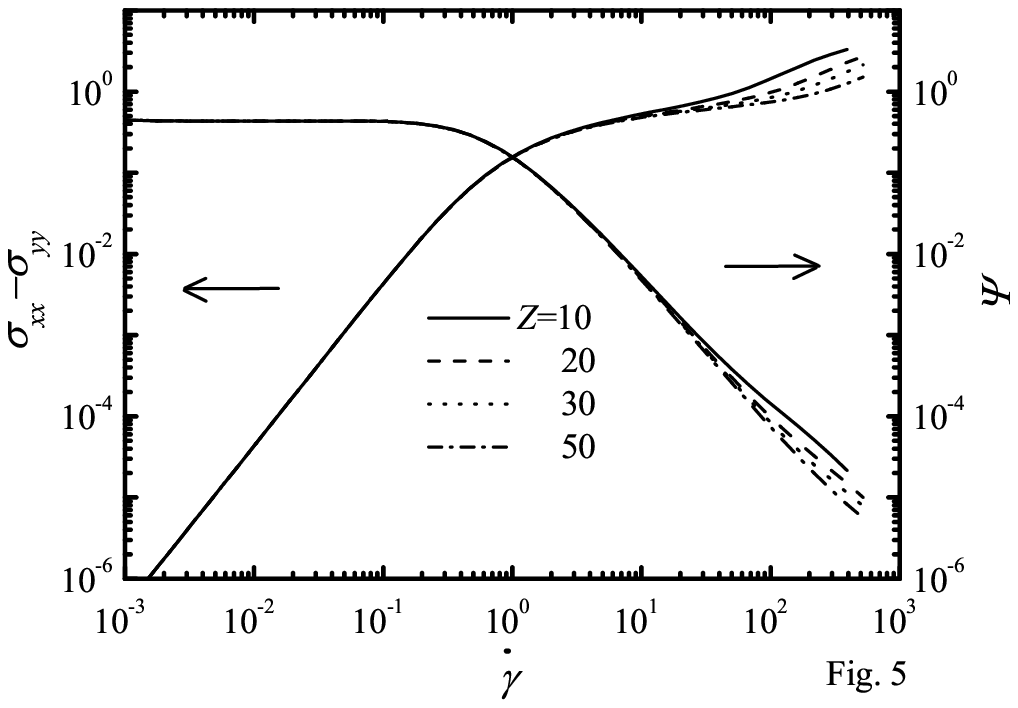}}
\label{05}
\end{figure}

\clearpage
\begin{figure}
\centerline{\includegraphics[angle=0,scale=1,draft=false]{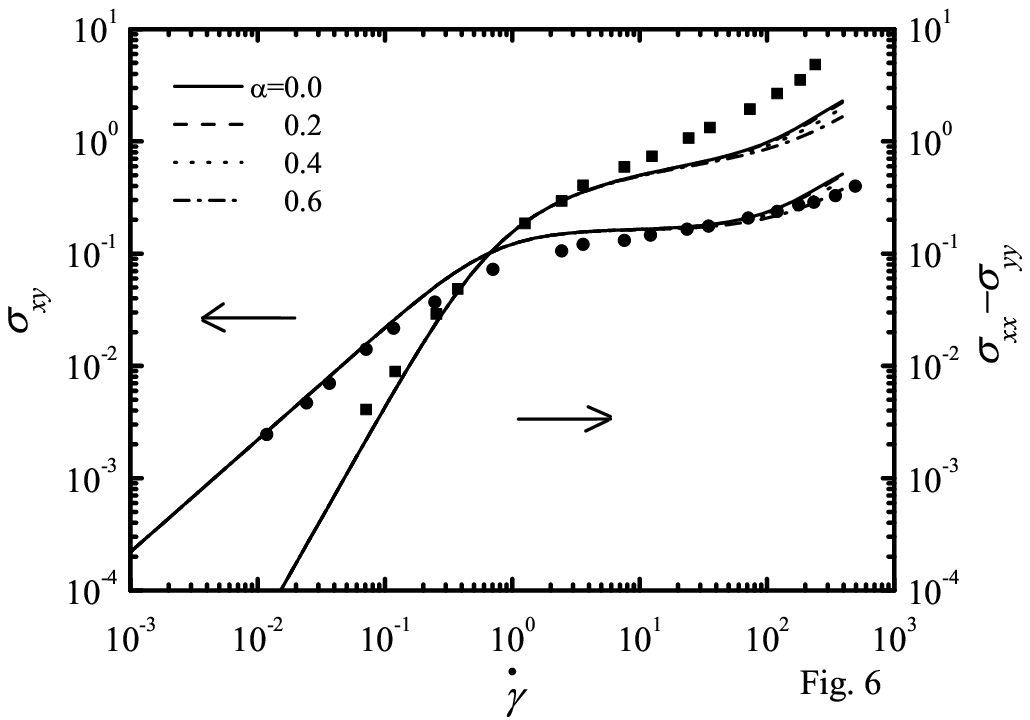}}
\label{06}
\end{figure}

\clearpage
\begin{figure}
\centerline{\includegraphics[angle=0,scale=1,draft=false]{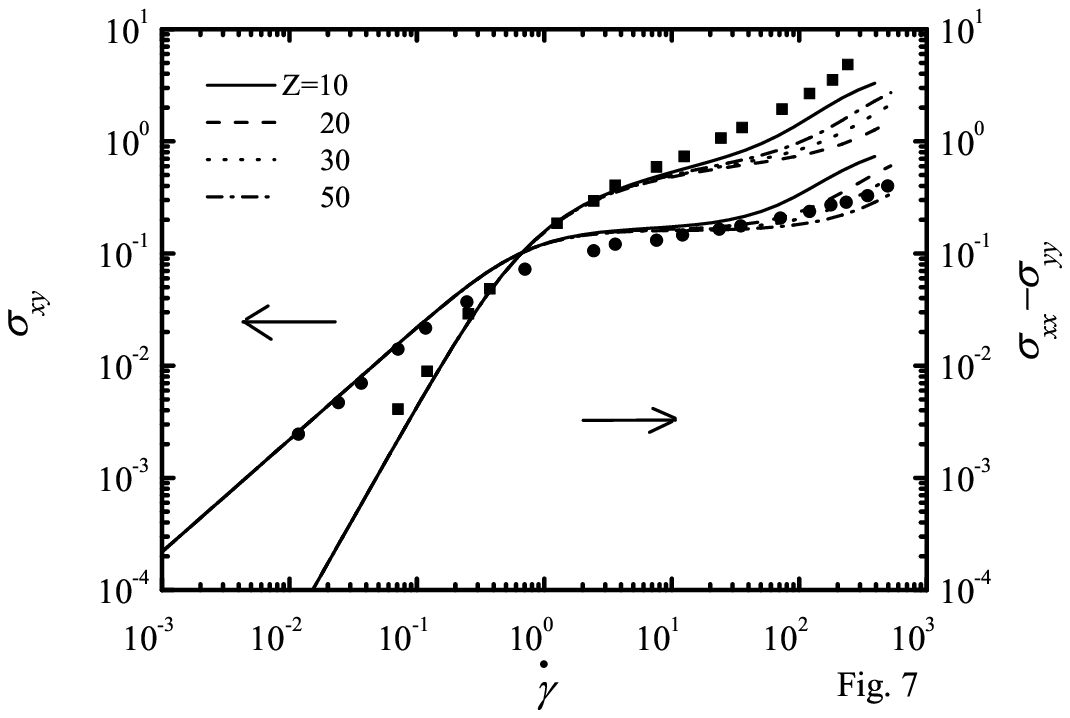}}
\label{07}
\end{figure}

\clearpage
\begin{figure}
\centerline{\includegraphics[angle=0,scale=1,draft=false]{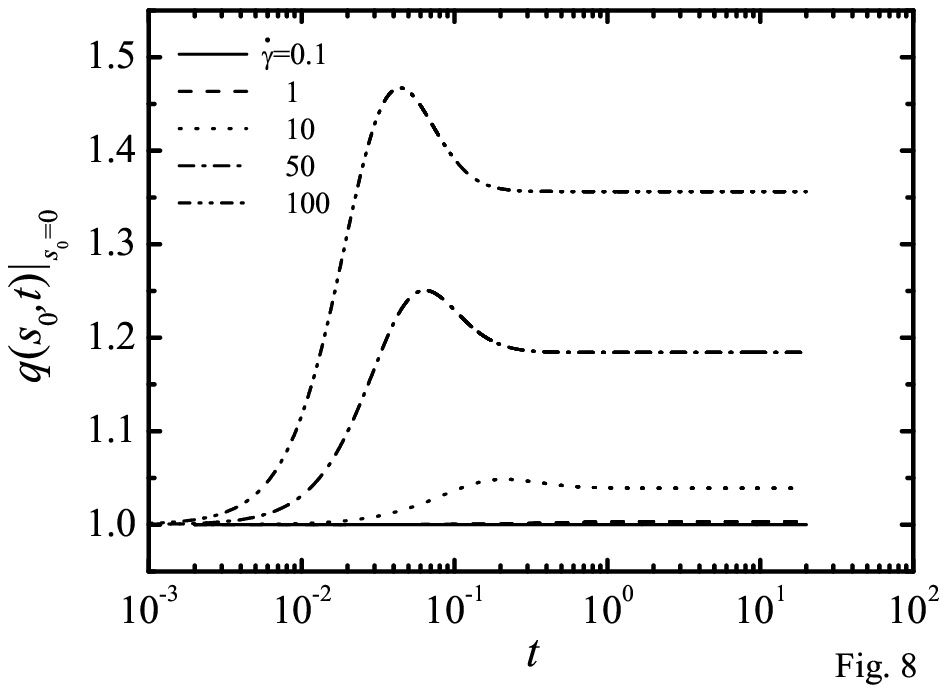}}
\label{08}
\end{figure}

\clearpage
\begin{figure}
\centerline{\includegraphics[angle=0,scale=1,draft=false]{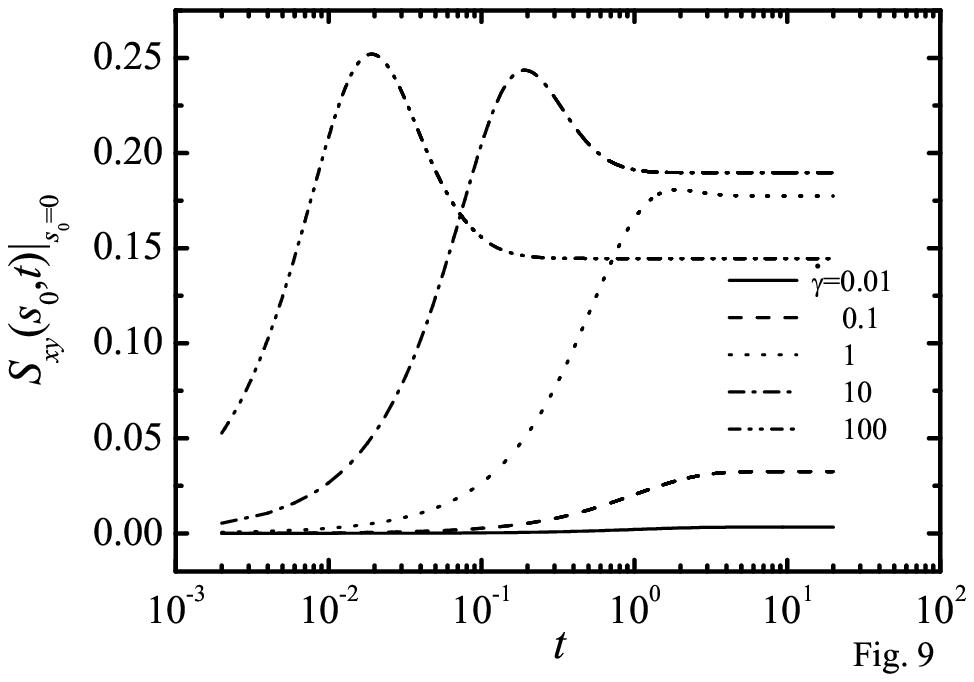}}
\label{09}
\end{figure}

\clearpage
\begin{figure}
\centerline{\includegraphics[angle=0,scale=1,draft=false]{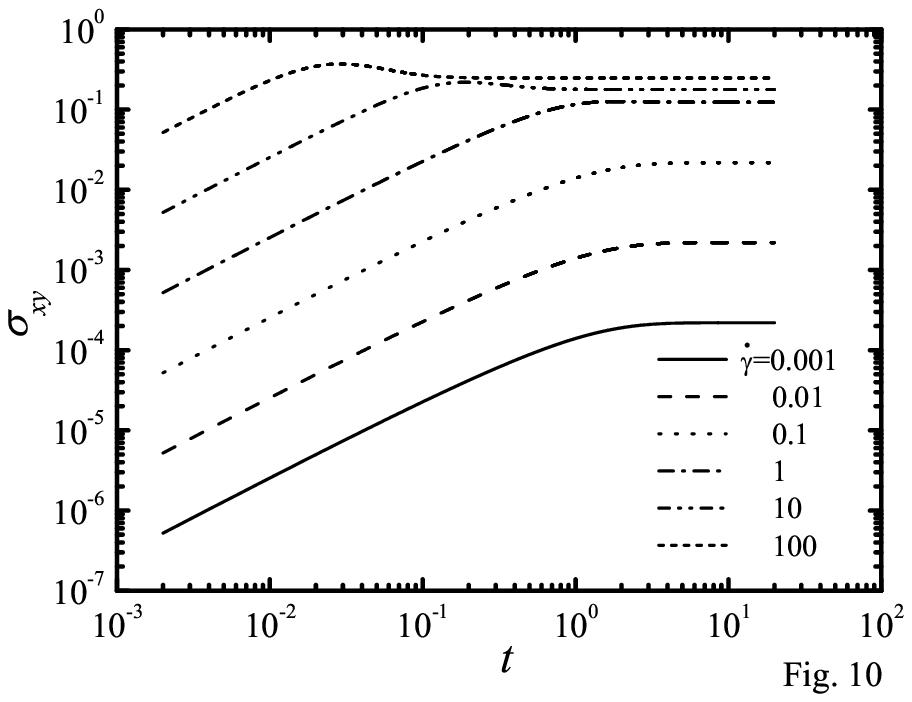}}
\label{10}
\end{figure}

\clearpage
\begin{figure}
\centerline{\includegraphics[angle=0,scale=1,draft=false]{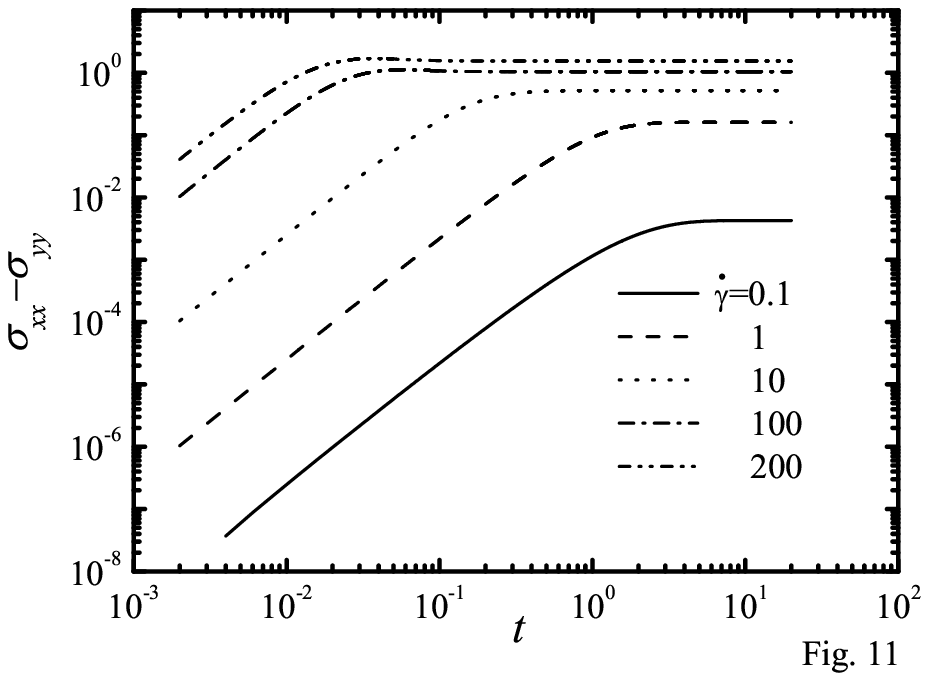}}
\label{11}
\end{figure}

\clearpage
\begin{figure}
\centerline{\includegraphics[angle=0,scale=1,draft=false]{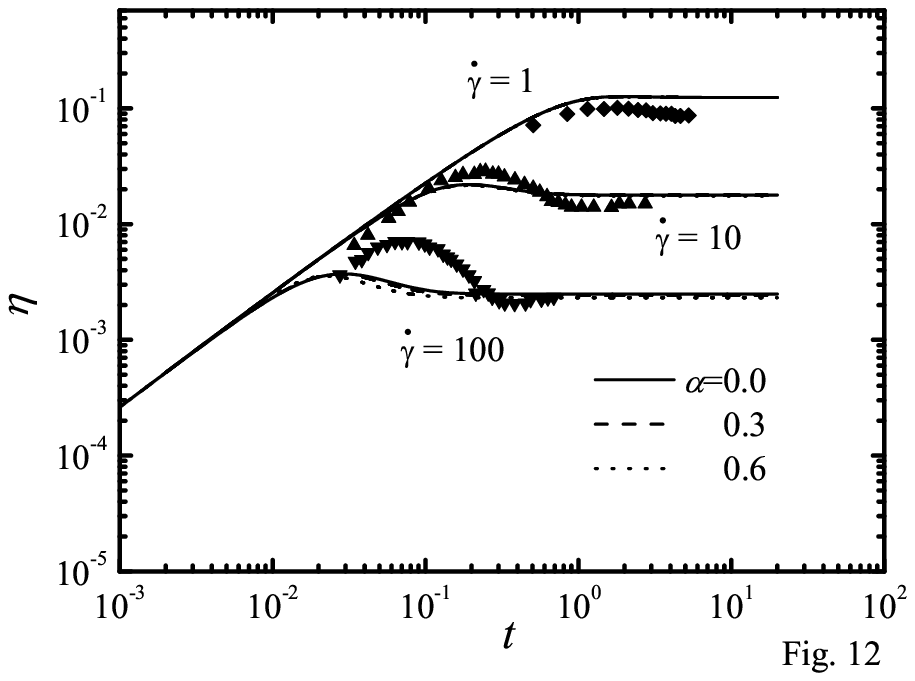}}
\label{12}
\end{figure}

\clearpage
\begin{figure}
\centerline{\includegraphics[angle=0,scale=1,draft=false]{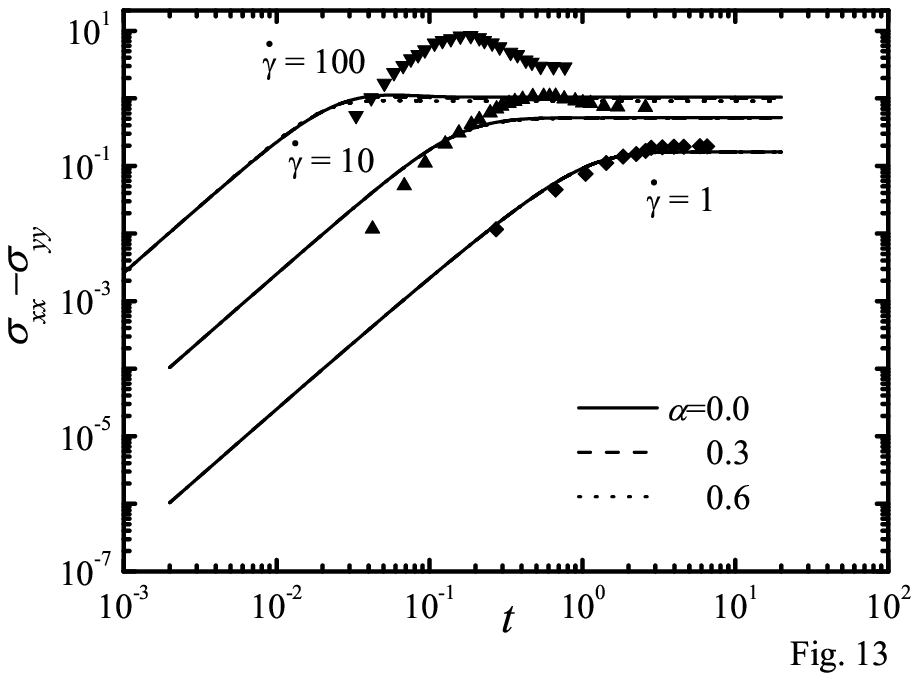}}
\label{13}
\end{figure}

\end{document}